\documentclass[graybox]{svmult}

\usepackage{mathptmx}       
\usepackage{helvet}         
\usepackage{courier}        
\usepackage{makeidx}         
\usepackage{graphicx}        
\usepackage{multicol}        
\usepackage[bottom]{footmisc}

\newcommand\be{\begin{equation}}
\newcommand\ee{\end{equation}}
\newcommand{\bea}{\begin{eqnarray}}
\newcommand{\eea}{\end{eqnarray}}

\newcommand{\nn}{\nonumber}

\def\p{\partial}

\def\a{\alpha}
\def\b{\beta}

\def\e{\epsilon}

\def\l{\lambda}

\def\vf{\varphi}
\def\G{\Gamma}
\def\D{\Delta}

\def\O{\Omega}

\def\ov{\over}

\def\[{\left[}
\def\]{\right]}
\def\({\left(}
\def\){\right)}
\def\<{\left\langle\,}
\def\>{\,\right\rangle}

\def\np#1#2#3{{Nucl. Phys.} {\bf B#1} (#2) #3}
\def\pl#1#2#3{{Phys. Lett. }{\bf B#1} (#2) #3}

\def\cmp#1#2#3{{Comm. Math. Phys.} {\bf #1} (#2) #3}

\def\tphi{\tilde\phi}
\def\hs{\hat{su}(2)}

\def\hepth#1{{arXiv:hep-th/}#1}

\def\np#1#2#3{{Nucl. Phys.} {\bf B#1} (#2) #3}
\def\pl#1#2#3{{Phys. Lett. }{\bf B#1} (#2) #3}

\def\cmp#1#2#3{{Comm. Math. Phys.} {\bf #1} (#2) #3}

\def\jhep#1#2#3{{JHEP} {\bf #1} (#2) #3}

\def\CL{{\cal L}}

\begin{document}

\title*{Renormalization in Some 2D $\hat{su}(2)$ Coset Models}
\author{Marian Stanishkov}
\institute{Marian Stanishkov \at Institute for Nuclear Research and Nuclear Energy, Bulgarian Academy of Sciences, Sofia 1784, Bulgaria, \email{marian@inrne.bas.bg}}
%
%
\maketitle
\abstract{We consider a RG flow in certain 2D coset models perturbed by the least relevant field. In the case of the symmetric su(2) coset model we show, up to second order of the perturbation theory, that there exists a nontrivial IR fixed point. We obtain the structure constants and the four-point functions of certain fields by deriving specific recursive relations. This allows us to compute the anomalous dimensions and the mixing coefficients of these fields in the UV and IR theories. In the case of another su(2) coset model, describing the N=2 superconformal theories, we show that there does not exists a nontrivial IR fixed point up to second order.}

\section{Introduction}
\label{sec:1}

In the first part of this paper we consider the symmetric $\hat{su}(2)$ coset model $M(k,l)$ \cite{gko} perturbed by the least relevant operator. It is known \cite{myo} that there exists an infrared fixed point of the renormalization group flow of this theory which coincides with the model $M(k-l,l)$.  Here we are interested in the mixing of certain fields under the corresponding RG flow. It is known that the mixing coefficients coincide for $l=1$ (Virasoro) and $l=2$ (superconformal) theories (particular cases of $M(k,l)$)
up to the second order of the perturbation theory \cite{as}. We will show that this is the case in the general theory, i.e. they do not depend on $l$ and are finite up to the second order.
For that purpose one needs in addition to the structure constants also the corresponding four-point functions which are not known exactly. We find it convenient, following \cite{myo}, to use the construction presented in \cite{myt}. Namely, we define the perturbing field and the other fields in consideration recursively as a product of lower level fields. Then the corresponding structure constants and four-point functions at some level $l$, governing the perturbation expansion, can be obtained recursively from those of the lower levels by certain projected tensor product.

In the second part of the paper we discuss the renormalization group properties of the $N=2$ superconformal minimal models. It is known that these models are connected to another $\hs$ based coset theories. The latter determine the so called parafermionic construction \cite{zfpf,zfpf1}. It is very useful for the calculation of the 4-point functions and the structure constants of the 2D OPE algebra. The reason for that is in the relation of the parafermionic models with the $su(2)$ Wess-Zumino-Witten (WZW) models \cite{wzwit}.
We compute the $\beta$ function up to second order in the perturbation theory and show that it doesn't possesses a non-trivial fixed point. We argue that this is true also in higher orders.

\section{Symmetric $\hs$ coset models}

In this Section we present the general $\hs$ coset model perturbed by the least relevant field. We obtain the $\beta$ function and show that it has a non-trivial fixed point up to  second order in the perturbation theory. We also construct certain fields and find their anomalous dimensions and the corresponding mixing matrix.

\subsection{The theory}

Consider a two-dimensional CFT $M(k,l)$ based on the coset:
\be\nn
{\hat{su}(2)_k\times \hat{su}(2)_l\over \hat{su}(2)_{k+l}},
\ee
where $k$ and $l$ are integers, we assume $k>l$. It is written in terms of $\hat{su}(2)_k$ WZNW models with current $J^a$, $k$ is the level. The latter are CFT's with a stress tensor expressed through the currents by the Sugawara construction, the central charge is $c_k={3k\ov k+2}$. The energy momentum tensor of the coset is then $T=T_k+T_l-T_{k+l}$ and:
$$
c={3kl(k+l+4)\over (k+2)(l+2)(k+l+2)}={3l\over l+2}\left(1-{2(l+2)\over (k+2)(k+l+2)}\right).
$$
The dimensions of the primary fields $\phi_{m,n}(l,p)$ of the "minimal models" ($m,n$ are integers) are computed in \cite{kmq}:
\bea\label{dmn}
\D_{m,n}(l,p) &=&{((p+l)m-p n)^2-l^2\over 4lp(p+l)}+{s(l-s)\over 2l(l+2)},\\
\nn  &=&|m-n|( mod (l)),\hskip1cm 0\le s\le l,\\
\nn &1&\le m\le p-1, \hskip1cm 1\le n\le p+l-1
\eea
where we introduced ${\bf p=k+2}$ (note that we inverted $k$ and $l$  in the definition of the fields).

In this paper we will use a description of the theory
$M(k,l)$ presented in \cite{myt}. It was shown there that this theory is not
independent but can be built out of products  of theories of lower
levels. Schematically this can be written as a recursion:
\be\label{proj}
M(1,l-1)\times M(k,l)={\bf P}(M(k,1)\times M(k+1,l-1))
\ee
where ${\bf P}$ in the RHS is a specific projection. It allows the
multiplication of fields of the same internal indices and describes
primary and descendent fields.

In the following we will be interested in the CFT $M(k,l)$ perturbed by the least
relevant field.  The theory is described by the Lagrangian:
$$
\CL(x)=\CL_0(x)+\l \tilde\phi(x)
$$
where $\CL_0(x)$ describes the theory $M(k,l)$ itself. We define the field $\tilde\phi=\tilde\phi_{1,3}$ in terms of lower level fields:
\be\label{field}
\tilde\phi_{1,3}(l,p)=a(l,p)\phi_{1,1}(1,p)\tilde\phi_{1,3}(l-1,p+1)+b(l,p)\phi_{1,3}(1,p)\phi_{3,3}(l-1,p+1).
\ee
Here the field $\phi_{3,3}(l,p)$ is just a primary field form (\ref{dmn}). The dimension of the field (\ref{field}) is:
\be\label{delt}
\D=\D_{1,3}+{l\over l+2}=1-{2\over p+l}=1-\e.
\ee
In this paper we consider the case $p\rightarrow\infty$ and
assume that $\e={2\over p+l}\ll 1$ is a small parameter. The coefficients $a(l,p)$ and $b(l,p)$ as well as the structure constants of the fields involved in the construction (\ref{field}) can be found by demanding the closure of the fusion rules \cite{myo}.

The mixing of the fields along the RG flow is connected to the two-point function. Up to the second order of the perturbation theory it is given by:
\bea\label{twop}
<\phi_1(x)\phi_2(0)>&=&<\phi_1(x)\phi_2(0)>_0-\l\int <\phi_1(x)\phi_2(0)\tilde\phi(y)>_0 d^2y+\\
\nn &+&{\l^2\ov 2}\int <\phi_1(x)\phi_2(0)\tilde\phi(x_1)\tilde\phi(x_2)>_0 d^2x_1 d^2x_2 +\ldots
\eea
where $\phi_1$, $\phi_2$ can be arbitrary fields of dimensions $\D_1$, $\D_2$. The first order corrections are expressed through the structure constants. Let us focus here on the second order. One can use the conformal transformation properties of the fields to bring the double integral to the form:
\bea\label{bint}
&\int& <\phi_1(x)\phi_2(0)\tilde\phi(x_1)\tilde\phi(x_2)>_0 d^2x_1d^2x_2 =\\
\nn &=&(x\bar x)^{2-\D_1-\D_2-2\D}\int I(x_1) <\tilde\phi(x_1)\phi_1(1)\phi_2(0)\tilde\phi(\infty)>_0 d^2x_1
\eea
where:
\be\nn
I(x)=\int |y|^{2(a-1)}|1-y|^{2(b-1)}|x-y|^{2c} d^2y
\ee
and $a=2\e+\D_2-\D_1$,$b=2\e+\D_1-\D_2$, $c=-2\e$. It is well known that the integral for $I(x)$ can be expressed in terms of hypergeometric functions whose behaviour around the points $0$, $1$ and $\infty$ is well known.
It is clear that the integral (\ref{bint}) is singular.
We follow the
regularization procedure proposed in \cite{pogt} . It was proposed there to cut discs in the two-dimensional surface of radius $r$
(${1\ov r}$) around singular points $0$, $1$ ($\infty$) with $0\ll r_0\ll r<1$, where $r_0$ is the ultraviolet cut-off.
The additional parameter $r$ is not physical and should not appear in the
final result. The region outside these discs, where the integration is well-defined, is called $\O_{r,r_0}$.
Near the
singular points one can use the OPE. The final result is a sum of all these contributions. It turns out however that
we count twice two lens-like regions around the point $1$ so we have
to subtract those integrals.

Let us consider the correlation function that enters the integral (\ref{bint}). The basic ingredients for the computation of the four-point correlation functions are the conformal blocks. According to the construction (\ref{proj}) any field $\phi_{m,n}(l,p)$ (or its descendent) can be expressed recursively as a product of lower level fields. Therefore the corresponding conformal blocks will be a product of lower level conformal blocks. Due to the RHS of (\ref{proj}) only certain products of conformal blocks will survive the projection ${\bf P}$.

Let us consider for example the correlation function of the perturbing field itself. The corresponding conformal blocks are linear combinations of products of conformal blocks at levels $1$ and $l-1$. In view of the construction (\ref{field}) there are in general 16 terms. Some of them are absent because of the fusion rules in each intermediate channel. Here there are three channels: identity $\phi_{1,1}$, the field $\tilde\phi_{1,3}$ itself and the descendent field $\tilde\phi_{1,5}$ which is defined in a way similar to that of $\tilde\phi_{1,3}$. We compute the conformal blocks up to a sufficiently high level and make a guess (remind that we need the result in the leading order in $\e\rightarrow 0$).
As a result, we obtain the following 2D correlation function:
\bea\label{tpf}
&<&\tphi(x)\tphi(0)\tphi(1)\tphi(\infty)>=\\
\nn &=&\left|{1\ov x^2(1-x)^2}\left[1-2x+({5\ov 3}+{4\ov 3l})x^2-({2\ov 3}+{4\ov 3l})x^3+{1\ov 3}x^4\right]\right|^2+\\
\nn &+&{16\ov 3l^2}\left|{1\ov x(1-x)^2}\left[1-{3\ov 2}x+{l+1\ov 2}x^2-{l\ov 4}x^3\right]\right|^2+\\
\nn &+&{5\ov 9}\left({2(l-1)\ov l}\right)^2\left|{1\ov (1-x)^2}\left[1-x+{l\ov 2(l-1)}x^2\right]\right|^2.
\eea
One can check that this function is crossing symmetric and has a correct behaviour near the singular points.

We now use this function for the computation of the $\b$-function up to the second order. For that purpose we have to compute the integral in (\ref{bint}).
The integration over the safe region far from the singularities
yields ($I(x)\sim {\pi\ov \epsilon}$):
\bea\nn
&\int_{\Omega_{r,r_0}}& I(x)<\tilde\phi(x)\tilde\phi(0)\tilde\phi(1)\tilde\phi(\infty)>d^2x=\\
&=&{(29 l^2-128 l) \pi^2\ov 24 \e l^2} + {2 \pi^2\ov\e r^2} + {\pi^2\ov 2 \e r_0^2} -
{ 64 \pi^2 \log r\ov 3 \e l^2} - {32 \pi^2 \log 2 r_0\ov 3 \e l^2}
\eea
and we omitted the terms of order $r$ or $r_0/r$.

We have to subtract the integrals over the lens-like regions
since they  would be accounted twice.
Here is the result of that integration:
$$
{\pi^2 \ov \e}
\(-{1\ov r^2}+{1\ov 2 r_0^2} +{1\ov 24}(29+{64\ov l})+{32\ov 3 l^2}\log{r\ov 2r_0}\).
$$
Next we have to compute the integrals near the singular points $0,1$ and $\infty$. For that purpose we can use the OPE of the
fields and take the appropriate limit of $I(x)$.
Near the point $0$ the relevant OPE is (by definition (\ref{field}):
$$
\tilde\phi(x)\tilde\phi(0)=(x\bar x)^{-2\Delta}(1+\ldots)
+ C_{(1,3)(1,3)}^{(1,3)}(x\bar x)^{-\Delta}(\tilde\phi(0)+\ldots).
$$
The structure constant was computed in \cite{myo}.
The value of $I(x)$ near $0$ is given in \cite{pogt} and finally we obtain:
\be\nn
\int_{D_{r,0}\backslash D_{r_0,0}} I(x)<\tilde\phi(x)\tilde\phi(0)\tilde\phi(1)\tilde\phi(\infty)>d^2x
=-{\pi^2\ov r^2 \e} + {32 \pi^2\ov
 3 l^2 \e^2} -{32 \pi^2\ov l\e} + {32\ov 3l^2} {\pi^2 \log r\ov\e}
\ee
where the region of integration ${D_{r,0}\backslash D_{r_0,0}}$ is a ring with internal and external radiuses $r_0$ and $r$ respectively.
Since the integral near $1$ gives obviously the same result, we just need to add the above result twice.
To compute the integral near infinity, we use a relation
$$
<\phi_1(x)\phi_2(0)\phi_3(1)\phi_4(\infty)>=(x\bar x)^{-2\D_1}<\phi_1(1/x)\phi_4(0)\phi_3(1)\phi_2(\infty)>
$$
and $I(x)\sim {\pi\ov\e}(x\bar x)^{-2\e}$.
This gives
$$
\int_{D_{r,\infty}\backslash D_{r_0,\infty}} I(x)<\tphi(x)\tphi(0)\tphi(1)\tphi(\infty)>d^2x
=-{\pi^2\ov r^2 \e} + {16 \pi^2\ov 3 l^2\e^2}  - 16{ \pi^2\ov l\e} + {32 \pi^2 \log r\ov 3l^2\e}
$$
where now ${D_{r,\infty}\backslash D_{r_0,\infty}}$ is a ring between ${1\ov r}$ and ${1\ov r_0}$.

Putting altogether, we obtain the finite part of the integral:
\be\nn
{80\pi^2\ov 3l^2 \e^2}-{88\pi^2\ov l\e}.
\ee
We want to remind also that we follow the renormalization scheme
proposed in \cite{pogt}. Therefore we already omitted the terms
proportional to $r_0^{4\e-2}$ which could be canceled by an
appropriate counterterm in the action.

Taking into account also the first order term, we get the final
result (up to the second order) for the two-point function of the
perturbing field:
\bea\label{twopt}
G(x,\l)&=&<\tphi (x)\tphi(0)>\\
\nn &=&(x\bar x)^{-2+2\e}\left[1-\l {4\pi\ov \sqrt 3}\left({2\ov l\e}-3\right)(x\bar x)^\e+{\l^2\ov 2}\left({80\pi^2\ov 3l^2 \e^2}-{88\pi^2\ov l\e}\right)(x\bar x)^{2\e}
+\ldots\right].
\eea
We now introduce a renormalized coupling constant $g$ and a renormalized field $\tphi^g=\p_g {\cal L}$ analogously to $\tphi=\p_{\l}{\cal L}$.
It is normalized by $<\tphi^g (1)\tphi^g(0)>=1$.
In this renormalization scheme the $\b$-function is given by \cite{zam,pogt}:
$$
\beta(g)=\e\l{\p g \ov\p\l}=\e\l\sqrt{ G(1,\l)}
$$
One can invert this and compute the bare coupling constant and the $\beta$-function in terms of $g$:
\bea\label{bare}
\l&=&g+g^2{\pi\ov \sqrt 3}\left({2\ov l\e}-3\right)+g^3{\pi^2\ov 3}\left({4\ov l^2\e^2}-{10\ov l\e}\right)+{\cal O}(g^4),\\
\beta(g)&=&\e g-g^2{\pi\ov\sqrt 3}({2\ov l}-3\e)-{4\pi^2\ov 3l}g^3+{\cal O}(g^4).
\eea
A nontrivial IR fixed point occurs at the zero of the $\beta$-function:
\be\label{fx}
g^*={l\sqrt{3}\ov 2\pi}\e(1+{l\ov 2}\e).
\ee
It corresponds to the IR CFT  $M(k-l,l)$ as can be seen from the central charge difference:
$$
c^*-c=-{4(l+2)\ov l}\pi^2\int_0^{g^*}\beta(g)d g=-l(1+{l\ov 2})\e^3-{3l^2\ov 4}(l+2)\e^4+{\cal O}(\e^5).
$$
The anomalous dimension of the perturbing field becomes
$$
\D^*=1-\p_g\beta(g)|_{g^*}=1+\e+l\e^2+{\cal O}(\e^3)
$$
which matches with that of the field $\phi_{3,1}(l,p-l)$ of $M(k-l,l)$ (defined precisely below).

\subsection{Mixing of the fields}

Let us define recursively the descendant fields $\tilde\phi_{n,n\pm 2}$:
\bea\nn
\tilde\phi_{n,n+2}(l,p)&=&x\phi_{n,n}(1,p)\tilde\phi_{n,n+2}(l-1,p+1)+y\phi_{n,n+2}(1,p)\phi_{n+2,n+2}(l-1,p+1),\\
\nn \tilde\phi_{n,n-2}(l,p)&=&\tilde x\phi_{n,n}(1,p)\tilde\phi_{n,n-2}(l-1,p+1)+\tilde y\phi_{n,n-2}(1,p)\phi_{n-2,n-2}(l-1,p+1)
\eea
(where $x$, $\tilde x$ and $y$, $\tilde y$ are at $(l,p)$) and the derivative $\partial\phi_{n,n}$ of the primary field
\be\nn
\phi_{n,n}(l,p)=\phi_{n,n}(1,p)\phi_{n,n}(l-1,p+1).
\ee
They have dimensions close to $1$
\bea\label{dimn}
\tilde\D_{n,n\pm 2} &=&1+{n^2-1\ov 4p}-{(2\pm n)^2-1\ov 4(p+l)}=1-{1\pm n\ov 2}\e+O(\e^2),\\
\nn 1+\D_{n,n} &=&1+{n^2-1\ov 4p}-{n^2-1\ov 4(p+l)}=1+{(n^2-1)l\ov 16}\e^2+O(\e^3).
\eea
This suggests that they mix along the RG-trajectory. To ensure this we ask that their fusion rules with the perturbing field are closed. This requirement defines the coefficients and the corresponding structure constants \cite{myrg}.

We want to compute the matrix of anomalous dimensions and the corresponding mixing matrix of the fields defined above. For that purpose we compute their two-point functions up to second order and the corresponding integrals (\ref{bint}). The first order integrals are proportional to the structure constants. For the second order calculation we need the corresponding four point functions. They are obtained in a way similar to that of the perturbing field $\tphi(z)$ itself. The explicit form of the four-point functions we need: $<\tphi(x)\tphi(0)\tphi_{n,n+2}(1)\tphi_{n,n+2}(\infty)>$,\\ $<\tphi(x)\tphi(0)\tphi_{n,n+2}(1)\tphi_{n,n-2}(\infty)>$ and $<\tphi(x)\tphi(0)\tphi_{n,n}(1)\tphi_{n,n+2}(\infty)>$ can be found in \cite{mysec}.

Let us describe briefly the renormalization scheme.
We introduce renormalized fields $\phi^g_\a$ which are expressed through the bare ones by:
\be\label{defb}
\phi^g_\a=B_{\a\b}(\l)\phi_\b
\ee
(here $\phi$ could be a primary or a descendent field).
The two-point functions of the renormalized fields
\be\label{norm}
G_{\a\b}^g(x)=<\phi_\a^g(x)\phi_\b^g(0)>,\quad G_{\a\b}^g(1)=\delta_{\a\b}
\ee
satisfy the Callan-Symanzik equation:
$$
(x\p_x-\b(g)\p_g)G_{\a\b}^g+\sum_{\rho=1}^2(\G_{\a\rho}G_{\rho\b}^g+\G_{\b\rho}G_{\a\rho}^g)=0.
$$
The matrix of anomalous dimensions $\Gamma$ that appears above is given by
\be\label{ano}
 \G=B\hat\D B^{-1}-\e\l B\p_\l B^{-1}
 \ee
where $\hat\D=diag(\D_1,\D_2)$ is a
diagonal matrix of the bare dimensions.
The matrix $B$, as defined in (\ref{defb}), is
computed from the matrix of the bare two-point functions we computed,
using the normalization condition (\ref{norm}) and requiring the matrix
$\G$ to be symmetric.

Let us combine the fields in consideration in a vector with components:
$$
\phi_1=\tphi_{n,n+2},\quad
\phi_2=(2\D_{n,n}(2\D_{n,n}+1))^{-1}\p\bar\p \phi_{n,n},\quad
\phi_3=\tphi_{n,n-2}.
$$
The field $\phi_2$ is normalized so that its bare two-point function is $1$.

We can write the matrix of the bare two-point functions $G_{\a,\b}(x,\l)=<\phi_\a(x)\phi_\b(0)>$ up to the second
order in the perturbation expansion as:
\be\label{expan}
G_{\a,\b}(x,\l)=
(x\bar x)^{-\D_\a-\D_\b}\left[\delta_{\a,\b}-\l C^{(1)}_{\a,\b}(x\bar x)^{\e}+{\l^2\ov 2}C^{(2)}_{\a,\b}(x\bar x)^{2\e}+...\right].
\ee
As we already mentioned, the two-point functions in the first order are proportional to the
structure constants \cite{zam}.
The second order contribution is a result of the double integration in (\ref{bint}) of the four-point functions mentioned above. This integration goes along the same lines as in the case of the perturbing field.

Using the entries $C^{(1)}$ and $C^{(2)}$ thus obtained we can apply the renormalization procedure and obtain the matrix of anomalous dimensions (\ref{ano}). The bare coupling constant $\l$ is
expressed through $g$ by (\ref{bare}) and the bare dimensions, up to order $\e^2$. Evaluating this matrix at the fixed point (\ref{fx}), we get:
\bea\nn
\G_{1,1}^{g^*}&=&1 + {(20 - 4 n^2) \e\ov 8 (n+1)} + {l(39 - n - 7 n^2 + n^3) \e^2\ov
 16 (n+1)},\\
\nn \G_{1,2}^{g^*}&=&\G_{2,1}^{g^*}={(n-1) \sqrt{{n+2\ov n}} \e(1+l\e)\ov n+1},\\
\nn \G_{1,3}^{g^*}&=&\G_{3,1}^{g^*}=0,\\
\nn \G_{2,2}^{g^*}&=&1 + {4 \e\ov n^2-1} + {l(65 - 2 n^2 + n^4) \e^2\ov 16 (n^2-1)},\\
\nn \G_{2,3}^{g^*}&=&\G_{3,2}^{g^*}={\sqrt{{n-2\ov n}} (n+1) \e(1+l\e)\ov n-1},\\
\nn \G_{3,3}^{g^*}&=&1 + {(n^2-5) \e\ov 2 (n-1)} + {l(-39 - n + 7 n^2 + n^3) \e^2\ov
 16 (n-1)}
\eea
Its eigenvalues are (up to order $\e^2$):
\bea\nn
\D_1^{g^*}&=&1 +  {1 + n\ov 2} \e + {l(7 +8 n + n^2)\ov 16} \e^2,\\
\nn \D_2^{g^*}&=&1 + {l(n^2-1)\ov 16} \e^2,\\
\nn \D_3^{g^*}&=&1 + {1-n\ov 2} \e +  {l(7 - 8 n + n^2)\ov 16} \e^2.
\eea
This result coincides with the dimensions $\tilde\D_{n+2,n}(l,p-l)$, $\D_{n,n}(l,p-l)+1$ and $\tilde\D_{n-2,n}(l,p-l)$ of the model $M(k-l,l)$ up to this order.
The corresponding normalized eigenvectors should be identified with the fields of $M(k-l,l)$:
\bea\nn
\tphi_{n+2,n}(l,p-l)&=&{2 \ov n (n+1)}\phi_1^{g^*} + {2
\sqrt{{n+2\ov n}}\ov n+1}\phi_2^{g^*} + {\sqrt{n^2-4}\ov
n}\phi_3^{g^*},\\
\nn \phi_2(l,p-l)&=&-{2 \sqrt{{n+2\ov n}}\ov n +
1}\phi_1^{g^*} -{n^2-5\ov n^2+1}\phi_2^{g^*} +{2\sqrt{{n-2\ov n}}\ov n-1}\phi_3^{g^*},\\
\nn \tphi_{n-2,n}(l,p-l)&=&{\sqrt{n^2-4}\ov n}\phi_1^{g^*}  - { 2
\sqrt{{n-2\ov n}}\ov n-1}\phi_2^{g^*} +{ 2\ov n(n-1)}\phi_3^{g^*}.
\eea
We used as before the notation $\tphi$ for the descendent field defined as in the UV theory and:
$$
\phi_2(l,p-l)={1\ov 2\D_{n,n}^{p-l}(2\D_{n,n}^{p-l}+1)}\p\bar\p
\phi_{n,n}(l,p-l)
$$
is the normalized derivative of the corresponding primary field. We notice that these eigenvectors are finite
as $\e\rightarrow 0$ with exactly the same entries as in $l=1$ \cite{pogt} and  $l=2$ \cite{as} minimal models. This is one of the main results of this paper.

\section{$N=2$ superconformal models}

The $N=2$ superconformal theories are invariant under the corresponding algebra generated by the stress-energy tensor $T(z)$, the supercurrents $G^{(\pm)}(z)$ and the $U(1)$ current $J(z)$. We shall be interested here in the simplest minimal models of this theory, labeled by an integer $p$, containing a finite number of fields. It is well known that the latter are connected to a coset
${\hs\times u(1)\over u(1)}$. The fields of the $N=2$ theories belong to different sectors, depending on the boundary conditions of the supercurrents. Here we will be interested in the fields of the Neveu-Schwartz (NS) sector only.

As it is clear from the coset construction, the $N=2$ superconformal minimal models admit a representation in terms of the  $D_{2p}$ parafermionic (PF) theories. It is based on the observation \cite{zfpf,zfpf1} of the fact that the generators of the $N=2$ supersymmetric theory could be expressed in terms of the PF currents and a free scalar field.

The primary fields in the $N=2$ theories are constructed from the lowest fields of the PF theory and exponentials of the free scalar field $\vf$. For the NS sector we have:
\bea\label{pfns}
N_m^l(z)&=&\phi_m^l(z)\exp{\left(i{m\ov \sqrt{2p(p+2)}}\vf(z)\right)},\\
\nn l&=&0,1,\ldots,p\quad m=-l,-l+2,\ldots,l,
\eea
where $\phi_m^l$ is the lowest dimensional fields of the parafermionic theory.

The $U(1)$ charge of this field is:
\be\nn
q_m^l={m\ov 2(p+2)}
\ee
and its dimension is simply the sum of the dimensions of the two ingredients:
\be\nn
\D_m^l=d_m^l+{m^2\ov 2p(p+2)}={l(l+2)\ov 4(p+2)}-{m^2\ov 4(p+2)}.
\ee
The product with the supercurrents defines the second component of the field $N_m^l$:
\be\nn
(N_m^l)^{II\pm}\sim \phi_{m\pm (p+2)}^{p-l}e^{i\left(m\pm(p+2)/\sqrt{2p(p+2)}\right)\vf}
\ee
Investigating the FR's in the NS sector  one must keep attention that they have more complicated structure due to the fact that there exist three different 3-point functions of the NS superfields - one even and two odd ones. The meaning of the odd FR's in terms of component fields  is that in the product of two first components of given superfields the second component of the RHS superfield appears. Taking all this into account we obtain the following FR's in the NS sector:
\bea\label{nsfr}
N^{l_1}_{m_1}N^{l_2}_{m_2}&=&\sum_{l=|l_1-l_2|}^L [\Psi^l_m],\\
\nn L&=&\min{(l_1+l_2,2p-l_1-l_2)}
\eea
where:
\bea\nn
\Psi^l_m&=&(N^l_{m_1+m_2})^{even},\quad |m_1+m_2|\le l,\\
\nn \Psi^l_m&=&(N^{p-l}_{m_1+m_2\pm (p+2)})^{odd},\quad |m_1+m_2|>l.
\eea

 In this Section we would like to discuss the renormalization group properties of the $N=2$ minimal models. In other words we would like to describe the RG flow of these models perturbed by the least relevant field. In the case of  $N=2$ minimal models the latter is constructed from the chiral and antichiral fields $N^p_{\pm p}$ of dimension $\D=1/2-1/(p+2)$ and $U(1)$ charge $q=\pm\D$. The suitable perturbation term, neutral and of dimension close to one, is therefore constructed out of the second components of such chiral fields. Explicitly we consider:
\be\nn
\CL=\CL_0+\int d^2z\Phi(z)
\ee
where $\CL_0$ represents the minimal model itself and the field $\Phi(z)$ is a combination of the second components:
\be\nn
\Phi=(N^p_p)^{II}+(N^p_{-p})^{II}\equiv \phi_++\phi_-.
\ee
It is neutral and has a dimension $\D=1-1/(p+2)=1-\e$. Similarly to what we did in the previous Section, we consider the case $p\rightarrow\infty$ and assume $\e=1/(p+2)$ to be a small parameter. Also, according to our parafermionic construction, we can express the perturbing field in terms of the PF currents and exponents of the scalar field as follows:
\bea\nn
(N^p_p)^{II}&=&\sqrt{{2p\ov p+2}}{\psi_1}^\dagger e^{-i{2\ov \sqrt{2p(p+2)}}}\equiv \phi_+,\\
\nn (N^p_{-p})^{II}&=&\sqrt{{2p\ov p+2}}{\psi_1} e^{i{2\ov \sqrt{2p(p+2)}}}\equiv \phi_-
\eea
where $\psi_1^{(\dagger)}$ are the simplest parafermionic currents.

Our purpose now is to compute the beta-function of this theory and to check for an eventual fixed point. For that we need to compute the two-point function of the perturbing field up to a second order. The expansion was already written in (\ref{twop}). As in the case of the symmetric coset, we need the 3- and 4-point functions of the perturbing field. We note that, due to the FR's computed above, the 3-point function of the field $\Phi(z)$, and therefore the first term in (\ref{twop}), is identically zero. So we are left with the computation of the second order term only. This computation goes along the same lines as above. We need to compute the 4-point function of $\Phi(z)$ up to zeroth order in $\e$ and to integrate it in the safe region $\O_{r,r_0}$ far from the singularities. Near the singular points $0$, $1$ and $\infty$ we use the OPE's that we computed above.

The 4-point function of the perturbing field $\Phi(z)$ is expressed through the corresponding functions of the parafermionic fields which are known \cite{zfpf} and the trivial power-like contribution of the exponents. The final result is (up to zeroth order in $\e$):
\be\nn
<\Phi(x)\Phi(0)\Phi(1)\Phi(\infty)>=C|1+{1\ov x^2}+{1\ov (1-x)^2}|^2
\ee
where $C$ is some structure constant. We will not need its explicit expression here. The integration of this function over the safe region gives:
\be\nn
{2\pi^2\ov\e}\left({31\ov 16}+{1\ov r^2}+{1\ov 4r_0^2}\right).
\ee
From this we have to subtract the contribution of the lens-like region:
\be\nn
{\pi^2\ov\e}\left({31\ov 16}-{1\ov r^2}+{1\ov 2r_0^2}\right).
\ee
At the end, we add the result of the integration near the singular points:
\be\nn
2\left(-{\pi^2\ov r^2\e}\right)+{2\pi^2\ov \e}\left(-{1\ov 2r^2}+{1\ov 2r_0^2}\right)
\ee
corresponding to the integrals around $0$ (and $1$) and $\infty$ respectively. Summing all the contributions we get finally as a result:
\be\nn
{\pi^2\ov \e r_0^2}.
\ee
Two comments are in order. First, this result contains only the cut-off parameter and could be cancelled by adding an appropriate counterterm in the action. Second, the finite contribution is identically zero. This means that there is no contribution to the beta-function neither in the first nor in the second order. One can speculate that this is the case also in higher orders. This result leads us to the conclusion that there do not exits a nontrivial fixed point of the beta-function close to the UV one. If such a fixed point exists it should be due to some non-perturbative effects.

\bigskip

\begin{acknowledgement}
This work is supported in part by the NSFB grant DFNI T02/6 and the bilateral grant STC/Bulgaria-France 01/6.
\end{acknowledgement}
\end{document}